\documentclass[aps,pra,twocolumn,groupedaddress,showpacs]{revtex4}
\usepackage{graphicx}
\usepackage{dcolumn}
\usepackage{bm}

\begin{document}

\title{Relativistic almost local hidden-variable theory}

\author{Hrvoje Nikoli\'c}
\affiliation{Theoretical Physics Division, Rudjer Bo\v{s}kovi\'{c}
Institute,
P.O.B. 180, HR-10002 Zagreb, Croatia.}
\email{hrvoje@thphys.irb.hr}

\date{\today}

\begin{abstract}
A simple relativistic quantum hidden-variable theory of particle trajectories,
similar to the Bohm theory but without nonlocal forces between the particles, is proposed.
To provide compatibility with statistical predictions of quantum mechanics
one needs to assume the initial probability density $|\psi|^2$ of particle positions in spacetime,
which is the only source of nonlocality in the theory.
This demonstrates that the usual Bohm hidden-variable theory contains much more nonlocality
than required by the Bell theorem.
\end{abstract}


\maketitle

The Bell theorem \cite{bell} (as well as some other theorems \cite{ghz,hardy})
shows that quantum mechanics (QM) 
is, to a certain extent, a nonlocal theory. In particular, the theorem
implies than any hypothetic hidden-variable completion of QM must 
necessarily be nonlocal. This is particularly manifest in the 
Bohm \cite{bell,bohm,BMbook1} nonlocal hidden-variable
formulation of nonrelativistic QM, where acceleration or velocity
of one particle depends on instantaneous positions of all other particles.
Yet, the existence of other interpretations of QM, in which
such instantaneous influences do not play any role, suggests
that the Bohm interpretation might contain more
nonlocality than necessary. For example, if the wave function itself
satisfying the Schr\"odinger equation 
is the only objectively existing entity \cite{mw}, then the only
source of nonlocality is the nonseparability of the wave function, while all
equations of motion are local. The similar is true
for the standard instrumental view of QM,
which is agnostic on the issue of objective reality.
The fact that these ``non-Bohmian'' views of QM are also compatible with the Bell
theorem suggests that the Bohm interpretation might contain
much more nonlocality than required by the Bell theorem. 

Such a view is also supported by a recent
demonstration that the Bohm theory can be 
reformulated in an apparently local form \cite{norsen},
but in a very complicated way in terms of an infinite 
tower of auxiliary pilot waves in the physical
(rather than configuration) space, satisfying an infinite coupled set
of local equations of motion. However, the fact that
the set of equations is infinite leaves a space for
suspicions that it could still be a nonlocal theory in disguise.

To provide a more compelling argument 
that the usual Bohm formulation of QM contains much more nonlocality
than required by the Bell theorem,
in this paper we propose a simple hidden-variable theory of particle
trajectories very similar to the Bohm theory, but without 
nonlocal forces between the particles. Instead, with a given
wave function in the configuration space, the velocity of a particle
depends only on the position of {\em that} particle.
The compatibility with statistical predictions of QM is encoded
in the initial correlations between the particles, which are
nonlocal in accordance with the Bell theorem.
These {\em initial} correlations turn out to be the only
source of nonlocality in the theory.

Another distinguished feature of our theory is that it is explicitly
relativistic covariant. Time and space coordinates are treated
on an equal footing. In particular, the usual space probability density given
by $|\psi|^2$ is generalized to the spacetime probability density
(see, e.g., \cite{stuc,broyles} for old forms of that idea). It seems that
without such a relativistic probabilistic interpretation,
the compatibility between the local equations for particle trajectories and
probabilistic predictions of QM could not be achieved.

Let $x=\{x^{\mu}\}$, $\mu=0,1,2,3$, denotes the coordinates
of a position in spacetime.
The state of $n$ free (but possibly entangled!) relativistic spin-0 particles
can be described by a many-time wave function
$\psi(x_1,\ldots,x_n)$. This wave function satisfies $n$ Klein-Gordon equations
(with the units $\hbar=c=1$ and the Minkowski metric signature $(+,-,-,-)$)
\begin{equation}\label{KG}
(\partial_a^{\mu}\partial_{a\mu} +m_a^2)\psi(x_1,\ldots,x_n)=0,
\end{equation}
one for each $x_a$, $a=1,\ldots, n$.
(The Einstein convention of summation over repeated indices refers
only to vector indices $\mu$, not to particle labels $a$.) 
From this wave function one can construct the quantity
\begin{equation}\label{curn}
j_{\mu_1\ldots\mu_n}(x_1,\ldots,x_n) \equiv \left( \frac{i}{2} \right)^n \psi^*
\!\stackrel{\leftrightarrow}{\partial}_{\mu_1}\! \cdots
\!\stackrel{\leftrightarrow}{\partial}_{\mu_n}\! \psi ,
\end{equation}
where $\chi \!\stackrel{\leftrightarrow\;}{\partial_{\mu}}\! \varphi \equiv
\chi\partial_{\mu}\varphi - (\partial_{\mu}\chi)\varphi$ and
$\partial_{\mu_a}\equiv \partial/\partial x^{\mu_a}_a$.
The quantity (\ref{curn})
transforms as an $n$-vector \cite{witt}. Eq.~(\ref{KG})  implies that 
this quantity satisfies
the conservation equation $\partial_{\mu_1}j^{\mu_1\ldots\mu_n}=0$
and similar conservation equations with other $\partial_{\mu_a}$.
Thus we have $n$ conservation equations
\begin{equation}\label{consn}
 \partial_{\mu_a}j^{\mu_1\ldots \mu_a \ldots \mu_n}=0 ,
\end{equation}
one for each $x_a$.
Assuming that $\psi$ is a superposition of positive-frequency solutions to (\ref{KG}),
$\psi$ can be normalized such that the $n$-particle Klein-Gordon norm is equal to 1.
Explicitly, this means that
\begin{equation}\label{globconsn}
\int_{\Sigma_1} dS^{\mu_1}_1 \cdots \int_{\Sigma_n} dS^{\mu_n}_n \,
j_{\mu_1\ldots\mu_n} =1,
\end{equation}
where $\Sigma_a$ are arbitrary 3-dimensional spacelike hypersurfaces and
\begin{equation}
dS_a^{\mu_a}=d^3x_a |g_a^{(3)}|^{1/2} n^{\mu_a}
\end{equation}
is the covariant measure of the 3-volume on $\Sigma_a$.
Here $n^{\mu_a}$ is the unit future-oriented vector normal to $\Sigma_a$,
while $g_a^{(3)}$ is the determinant of the induced metric on $\Sigma_a$.
The conservation equations (\ref{consn}) imply that
the left-hand side of (\ref{globconsn})
does not depend on the choice of timelike hypersurfaces
$\Sigma_1,\dots,\Sigma_n$.

Now we introduce $n$ 1-particle currents
$j_{a\mu}(x_a)$ by omitting the integration over $dS^{\mu_a}_a$ in
(\ref{globconsn}). For example, for $a=1$,
\begin{equation}\label{eq6}
j_{1\mu}(x_1)=
\int_{\Sigma_2} dS^{\mu_2}_2 \cdots \int_{\Sigma_n} dS^{\mu_n}_n \,
j_{\mu \mu_2\ldots\mu_n}(x_1,\ldots,x_n) ,
\end{equation}
which does not depend on the choice of timelike hypersurfaces
$\Sigma_2,\dots,\Sigma_n$
and satisfies $\partial_{1\mu} j_1^{\mu}=0$.
This implies the conservation equation
\begin{equation}\label{cons2}
 \sum_{a=1}^{n} \partial_{a\mu}j_a^{\mu}(x_a)=0.
\end{equation}

Next we study the integral curves of the vector fields $j_a^{\mu}(x_a)$. 
These integral curves can be represented by functions $\tilde{X}^{\mu}_a(\tilde{s})$
satisfying {\em local} differential equations
\begin{equation}\label{param1}
 \frac{d\tilde{X}^{\mu}_a(\tilde{s})}{d\tilde{s}} = j_a^{\mu}(\tilde{X}_a(\tilde{s})) ,
\end{equation}
where $\tilde{s}$ is an auxiliary scalar parameter 
(a generalized proper time \cite{niksuperlum}) that parameterizes the curves.
However, the curves in spacetime do not depend on their parameterization.
In particular, even though the equations  (\ref{param1}) are local,
nonlocal parameterizations can also be introduced. For example,
along the integral curves the following parameterization-independent equalities are valid 
\begin{equation}\label{param2}
 \frac{dx^{\mu}_a}{dx^{\nu}_b} = \frac{j^{\mu}_a(x_a)}{j^{\nu}_b(x_b)}=
\frac{v^{\mu}_a(x_1,\ldots,x_n)}{v^{\nu}_b(x_1,\ldots,x_n)} ,
\end{equation}
where
\begin{equation}\label{eq10}
v^{\mu}_a(x_1,\ldots,x_n) \equiv \frac{j^{\mu}_a(x_a)}{|\psi(x_1,\ldots,x_n)|^2} .
\end{equation}
Thus we see that the integral curves
of $j_a^{\mu}(x_a)$ can also be parameterized by different functions
$X^{\mu}_a(s)$ satisfying nonlocal equations of motion
\begin{equation}\label{param3}
 \frac{dX^{\mu}_a(s)}{ds} = v_a^{\mu}(X_1(s),\ldots,X_n(s)) .
\end{equation}
The conservation equation (\ref{cons2}) now can be written as 
\begin{equation}\label{cons3}
 \sum_{a=1}^{n} \partial_{a\mu}(|\Psi|^2 v^{\mu}_a) =0,
\end{equation}
where
\begin{equation}\label{Nnorm}
 \Psi(x_1,\ldots,x_n)=\frac{\psi(x_1,\ldots,x_n)}{N^{1/2}} ,
\end{equation}
and $N$ is a normalization constant to be fixed later.
Since $\Psi(x_1,\ldots,x_n)$ does not have an explicit dependence on $s$,
(\ref{cons3}) can also be written as
\begin{equation}\label{cons4}
\frac{\partial |\Psi|^2}{\partial s} + \sum_{a=1}^{n} \partial_{a\mu}(|\Psi|^2 v^{\mu}_a) =0.
\end{equation}

In \cite{nikfol}, the integral curves of $j_a^{\mu}(x_a)$ have been used as an auxiliary
mathematical tool. Here, using the new result above that these curves can also be
viewed as integral curves of $v^{\mu}_a(x_1,\ldots,x_n)$, we propose a different
interpretation of these curves. We propose that these integral curves are the actual particle
trajectories. The compatibility with statistical predictions of the ``standard'' purely probabilistic
interpretation of QM is provided by Eq.~(\ref{cons4}), now interpreted
as the relativistic equivariance equation \cite{BMbook1,nikijqi,nikbosfer,hern,nikqft,niktorino,nikSmatr}.
Namely, if a statistical ensemble of particles has the probability distribution
(on the relativistic $4n$-dimensional configuration space) equal to 
\begin{equation}\label{distrib}
\rho(x_1,\ldots,x_n)=|\Psi(x_1,\ldots,x_n)|^2
\end{equation}
for some initial $s$, then the equivariance equation (\ref{cons4}) provides that
the ensemble will have the distribution (\ref{distrib}) for {\em any} $s$.
(The scalar parameter $s$ itself can be interpreted as a relativistic analogue
of the Newton absolute time \cite{niksuperlum,nikSmatr}.)
In this sense, the particle trajectories (\ref{param3}) are compatible
with the quantum-mechanical probability distribution (\ref{distrib}).

A few additional remarks are in order. First, the probabilistic interpretation 
(\ref{distrib}) implies that $N$ in (\ref{Nnorm}) should be fixed to
\begin{equation}
 N=\int d^4x_1\cdots d^4x_n |\psi(x_1,\ldots,x_n)|^2 .
\end{equation}
To avoid dealing with an infinite $N$, one can confine the whole physical system into
a large but finite 4-dimensional spacetime box. Mathematically more rigorous ways 
of dealing with wave functions that do not vanish at infinity also exist,
such as the rigged Hilbert space \cite{ballentine}.

Second, the spacetime probabilistic interpretation (\ref{distrib}), generalizing the usual 
space probabilistic interpretation of nonrelativistic QM, has also been
studied in older literature, such as \cite{stuc,broyles}. A detailed discussion of 
compatibility of such a generalized probabilistic interpretation
with the usual probabilistic interpretation is presented in \cite{niktorino}.
In particular, in \cite{niktorino} it is explained how particle trajectories 
obeying (\ref{cons4}) are compatible with {\em all} statistical predictions
of QM, not only with statistical predictions on particle positions.
(For example, even though (\ref{param1}) may lead to superluminal velocities,
a measured velocity cannot be superluminal \cite{niktorino}.)
The key insight is that all observations can be reduced
to observations of spacetime positions of some macroscopic pointer observables. 

Third, in the classical limit, the Klein-Gordon equations (\ref{KG}) reduce 
to the classical relativistic Hamilton-Jacobi equation \cite{niksuperlum}.
Consequently, by evaluating (\ref{eq6}) in that limit, it can be shown that  
(\ref{param1}) reduces to the classical relativistic equation of motion.

Fourth, it is straightforward to generalize the theory above to particles
interacting with a classical gravitational or electromagnetic background,
by replacing the derivatives $\partial_{\mu}$ with the appropriate
covariant derivatives. Again, it can be shown that
the resulting theory has the correct classical limit. 

Fifth, the theory can be generalized to particles with spin
and even to quantum field theory, by appropriate adaptation
of the formal developments presented in \cite{nikqft,nikSmatr}. 

To summarize, our main results can be reexpressed in the following way.
Eqs.~(\ref{param3}) are nonlocal Bohmian-like equations of motion,
compatible with statistical predictions of QM due to the equivariance equation
(\ref{cons4}). However, owing to the specific form of (\ref{eq10})
in which all nonlocality is carried by a common scalar factor universal to {\em all} components of
$v^{\mu}_a$, the nonlocality of (\ref{param3}) is only an
apparent nonlocality.
This apparent nonlocality can be explicitly eliminated by choosing a
different parameterization of the particle trajectories, which leads to
the manifestly local equations of motion (\ref{param1}).
In this sense, our theory of particle trajectories is much more local 
than the usual Bohmian formulation of QM. Yet, a certain nonlocal feature
is still present. To provide consistency with statistical predictions
of QM, one must assume that the {\it a priori} probabilities of initial
particle positions $X^{\mu}_a(0)$ are given by (\ref{distrib}).
Thus, all nonlocality can be ascribed to initial nonlocal correlations
between the particle spacetime positions.
The nonlocal forces between the particles turn out to be superfluous.

Even if the theory described above does not describe the true reality
behind QM, at the very least it provides an example which explicitly demonstrates that 
quantum reality may be much less nonlocal then suggested
by the usual form of the Bohm interpretation. 
We believe that it significantly enriches the
general understanding of nonlocality and relativity in QM.


This work was supported by the Ministry of Science of the
Republic of Croatia under Contract No.~098-0982930-2864.


\begin{thebibliography}{99}

\bibitem{bell}
J. S. Bell, {\it Speakable and Unspeakable in Quantum Mechanics}
(Cambridge University Press, Cambridge, 1987).

\bibitem{ghz}
D. M. Greenberger, M. Horne and A. Zeilinger,
in {\it Bell's Theorem, Quantum Theory, and Conceptions
of the Universe}, edited by M. Kafatos (Kluwer Academic, Dordrecht, 1989).

\bibitem{hardy}
L. Hardy, Phys. Rev. Lett. {\bf 68}, 2981 (1992). 

\bibitem{bohm}
D. Bohm, Phys. Rev. {\bf 85}, 166 (1952);  
D. Bohm, Phys. Rev. {\bf 85}, 180 (1952).

\bibitem{BMbook1}
D. Bohm and B. J. Hiley, {\it The Undivided Universe} (Routledge, London, 1993).

\bibitem{mw}
B. S. DeWitt and N. Graham (eds.), {\it The Many-Worlds Interpretation of
Quantum Mechanics} (Princeton University Press, New Jersey, 1973).

\bibitem{norsen}
T. Norsen, arXiv:0909.4553

\bibitem{stuc}
E. C. G. St\"uckelberg, Helv. Phys. Acta {\bf 14}, 322 (1941);
E. C. G. St\"uckelberg, Helv. Phys. Acta {\bf 14}, 588 (1941).
\bibitem{broyles}
A. A. Broyles, Phys. Rev. D {\bf 1}, 979 (1970).

\bibitem{witt}
B. S. DeWitt and R. W. Brehme, Ann. Phys. {\bf 9}, 220 (1960).

\bibitem{niksuperlum}
H. Nikoli\'c, arXiv:1006.1986.

\bibitem{nikfol}
H. Nikoli\'c, Int. J. Mod. Phys. A {\bf 22}, 6243 (2007).

\bibitem{nikbosfer}
H. Nikoli\'c, Found. Phys. {\bf 39}, 1109 (2009).
\bibitem{nikijqi}
H. Nikoli\'c, Int. J. Quant. Inf. {\bf 7}, 595 (2009).
\bibitem{hern}
S. Hernandez-Zapata and E. Hernandez-Zapata, Found. Phys. {\bf 40}, 532 (2010).
\bibitem{nikqft}
H. Nikoli\'c, Int. J. Mod. Phys. A {\bf 25}, 1477 (2010).
\bibitem{niktorino}
H. Nikoli\'c, arXiv:1002.3226, to appear in Int. J. Quant. Inf.
\bibitem{nikSmatr}
H. Nikoli\'c, arXiv:1007.4946.

\bibitem{ballentine}
L. E. Ballentine, {\it Quantum Mechanics: A Modern Development}
(World Scientific Publishing, Singapore, 2000).













\end{thebibliography}
\end{document}